\documentclass{article} 

\usepackage{aaspp4}
\usepackage{graphics}

\begin{document} 

%%%%%%%%%%%%%%%%% TITLE %%%%%%%%%%%%%%%%%%%%%%% 

\title{Can streamer blobs prevent the buildup of the interplanetary
magnetic field?} 

\author{M. K. van Aalst}
\affil{Sterrekundig Instituut, Universiteit Utrecht, Postbus 80 000, 
3508 TA Utrecht, Netherlands; aalst@phys.uu.nl}
 
\author{P. C. H. Martens\protect\footnote{As of January 1998: Department 
of Physics, Montana State University, P.O. Box 173840, Bozeman, MT 
59717-3840 U.S.A.}, A. J. C. Beli\"{e}n}
\affil{Solar System Division, ESA Space Science
Department at Goddard Space Flight Center, Greenbelt, 20771, MD, U.S.A.; 
pmartens@solar.stanford.edu, sbelien@esa.nascom.nasa.gov}

%%%%%%%%%%%%%%%%% ABSTRACT %%%%%%%%%%%%%%%%%%%%

\begin{abstract}

Coronal Mass Ejections continuously drag closed magnetic field lines 
away from the Sun, adding new flux to the interplanetary magnetic field 
(IMF).\@ We propose that the outward-moving blobs that have been 
observed in helmet streamers are evidence of ongoing, small-scale 
reconnection in streamer current sheets, which may play an important 
role in the prevention of an indefinite buildup of the IMF.\@ 
Reconnection between two open field lines from both sides of a streamer 
current sheet creates a new closed field line, which becomes part of the 
helmet, and a disconnected field line, which moves outward. The blobs 
are formed by plasma from the streamer that is swept up in the trough of 
the outward moving field line. We show that this mechanism is supported 
by observations from SOHO/LASCO.\@ Additionally, we propose a thorough 
statistical study to quantify the contribution of blob formation to the 
reduction of the IMF, and indicate how this mechanism may be verified by 
observations with SOHO/UVCS and the proposed NASA STEREO and ESA Polar 
Orbiter missions.
 
\end{abstract}


%%%%%%%%%%%%%%%%% INTRODUCTION %%%%%%%%%%%%%%%%%

\section{Introduction}

Using the LASCO C2 and C3 coronagraphs on the SOHO spacecraft 
\markcite{BEA95}(Brueckner et al.\ 1995), Sheeley et al.\ 
\markcite{SEA97}(1997) discovered blobs of material moving outward in 
coronal helmet streamers. Helmet streamers consist of a bubble-like or 
arch-like wide body (the helmet), consisting of closed magnetic field. 
Above the cusp, the pointed top of the helmet, lies a current sheet 
which extends radially outward, separating the two directions of open 
magnetic field around the helmet (see Fig.~1, top panel). 
We adopt the view that this streamer structure is formed by 
reconnection of open field lines in the current sheet that is left 
behind by a CME, as suggested by, e.g., Kopp \markcite{K92}(1992) and 
Kahler \& Hundhausen \markcite{KH92}(1992). Sheeley et al.\ 
\markcite{SEA97}(1997) observed that the blobs originate in the streamer 
current sheet, right above the cusp. They have an initial size of about 
$1 \, R_{\sun}$ in the radial and $0.1 \, R_{\sun}$ in the transverse 
direction. The blobs move radially outward along the streamer, with 
increasing velocities from about $150 \, \rm kms^{-1}$ near $5 \, 
R_{\sun}$ to $300 \, \rm kms^{-1}$ near $25 \, R_{\sun}$. Because of 
their relatively small initial sizes, low intensities ({$\Delta I/{I} 
\lesssim 0.1$}), radial motions, slow but increasing velocities, and 
location in the streamer belt, Sheeley et al.\ \markcite{SEA97}(1997) 
conclude that these features passively trace the solar wind. Wang et 
al.\ \markcite{WEA98}(1998) carried out some more detailed observations 
of the spatial distribution, relative intensities, and shapes of the 
blobs, and observed a rather steady occurrence rate of about four blobs 
per day. We believe that the creation of these blobs is connected to a 
longstanding problem in solar physics: maintaining a roughly constant 
amount of flux in the interplanetary magnetic field (IMF).

Coronal Mass Ejections (CMEs) generally originate in regions of closed 
magnetic field. These field lines are torn away from the Sun and 
stretched out into distended loops that extend well into interplanetary 
space \markcite{H98}(Hundhausen 1997). In principle, each consecutive 
CME thus introduces new flux into the heliosphere. This would result in 
an indefinite growth of the IMF, which is not observed (e.g.\ 
\markcite{G62}Gold 1962; \markcite{G75}Gosling 1975; 
\markcite{MacQ80}MacQueen 1980; \markcite{McCGP92}McComas, Gosling, \& 
Phillips 1992). Apparently, some kind of reconnection takes place on the 
field lines that are torn away from the Sun, disconnecting the tops of 
the loops and returning new closed field lines to the Sun.   
 
The reconnection must start shortly after the CME's departure, but it 
cannot be restricted to that first period. In situ measurements near and 
beyond 1 AU (e.g.\ \markcite{GBH95}Gosling, Birn, \& Hesse 1995; 
\markcite{G96}Gosling 1996) show that most field lines within CMEs are 
still connected to the Sun at both ends. However, a few field lines are 
connected on only one end, while others are completely disconnected.\ 
Additionally, about one third of the CMEs exhibit a flux-rope topology, 
characterized by a series of helical field lines wrapped around a 
central axis. Both this helicity and the disconnected field lines can 
only be explained by magnetic reconnection behind the CME during the 
days before it reaches 1 AU (e.g.\ Gosling et al.\ \markcite{GBH95}1995). 
On the other hand, to prevent an indefinite buildup of the IMF by field 
lines that are still connected to the Sun, reconnection must also keep 
taking place after the CME has passed 1 AU, several days after it 
has left the Sun. 

Three types of events show evidence for reconnection directly behind a 
departing CME.\@ First, about one third of all CMEs are accompanied by 
long-duration (many hours) X-ray emission from expanding loops or 
arcades of loops (see \markcite{G93}Gosling 1993). Non-thermal emission 
at X-ray and radio wavelengths is often observed in conjunction with 
these events \markcite{WC95}(Webb \& Cliver 1995). A beautiful example 
of a reforming helmet streamer is presented by Hiei, Hundhausen, \& Sime 
\markcite{HHS93}(1993). All these observations confirm the model of Kopp 
\& Pneuman \markcite{KP76}(1976) which explains hot loop formation by 
magnetic reconnection behind the CME.\@ Second, there are some 
observations of moving metric type IV radio events which are interpreted as 
emission from electrons that have become trapped in disconnected 
plasmoids within CMEs \markcite{KEA89}(e.g.\ Kundu et al.\ 1989). Third, 
there are many coronagraph observations of large circular, ovoidal or 
outward U- or V-shaped structures that are usually interpreted as 
disconnected CMEs (e.g.\ \markcite{IH83}Illing and Hundhausen 1983). 
Webb \& Cliver \markcite{WC95}(1995), analyzing all space-borne 
coronagraph and eclipse data up to then, concluded that possibly over 10 
\% of all CMEs fall in this category. Unfortunately, due to their 
infrequency compared to CMEs, even these three mechanisms combined 
cannot explain the necessary amount of flux-disconnection.

An even bigger problem is posed by reconnection long after the CME.\@ 
There have been suggestions that such reconnection would occur in 
interplanetary space itself (e.g.\ \markcite{W71}Wilcox 1971),
but the new closed loops that would have to return to the Sun have not 
been observed (\markcite{G75}Gosling 1975; \markcite{McEA89}McComas et 
al.\ 1989). McComas et al.\ (\markcite{McEA89}1989, 
\markcite{McCEA91}1991) suggested that the large U-shaped disconnections 
described above are not related to a departing CME, but occur all by 
themselves across the heliospheric current sheet. However, triggering 
these large reconnections by emerging flux elsewhere in the corona seems 
rather ad-hoc \markcite{WC95}(Webb \& Cliver 1995). Additionally, these 
events are still quite infrequent compared to CMEs. Evidently, the 
mechanism, or combination of mechanisms, that can disconnect enough flux 
to offset the constant flow of CMEs has not yet been identified. 

\newpage

In this Letter, we suggest that the creation of blobs in coronal 
streamers is a mechanism for ongoing, small-scale reconnection that can 
disconnect open field lines and reform closed magnetic loops on the Sun.
Especially since the blobs occur not just shortly after CMEs, but 
also rather steadily under quiet coronal conditions, they may play an 
important role in maintaining the roughly constant IMF.\@ 

%%%%%%%%%%%%%%%% MECHANISM %%%%%%%%%%%%%%%%%%%%%

\section{Mechanism}

We suggest that the blobs are the result of reconnection between two 
open field lines from both sides of a streamer current sheet. 
Naturally, we speak about field lines only to visualize what is 
going on; in reality, a finite amount of magnetic field will be 
involved in such a reconnection event. The field line topologies before 
and after the reconnection are outlined in Fig.~1. 
Reconnection between the two innermost open field lines in the top panel 
creates two new field lines. The first, with both ends connected to the 
Sun, is a new closed loop which becomes part of the helmet (as in e.g.\ 
\markcite{KH92}Kahler \& Hundhausen 1992). The other one, with its ends 
extending into interplanetary space, will move outward. The blobs are 
formed by plasma from the streamer that is swept up in the trough of this 
outward moving field line (middle and bottom panels). Considering the 
exponential decay of the density in the streamer with radial distance, 
almost all of the material in the blob is collected right away, just 
above the cusp. 

The loop that is moving outward will at first attempt to do so at the 
Alfv\'{e}n speed. However, it is immediately slowed down when it sweeps
up the plasma on its way. Nevertheless, the blob of material in the 
field line will accelerate faster than the surrounding wind, at least 
until the Lorentz force is balanced by the pressure differential that is 
built up when the plasma is swept up into a blob or, alternatively, 
until $\beta \gtrsim 1$ and the surrounding gas pressure starts to 
dominate the magnetic field. From then on, the field lines, and 
therefore the blobs, more or less flow along with their surroundings, 
tracing, as Sheeley et al.\ \markcite{SEA97}(1997) suggested, the slow 
solar wind.

\vspace*{3 mm}

\begin{center}
\scalebox{0.4}{\includegraphics{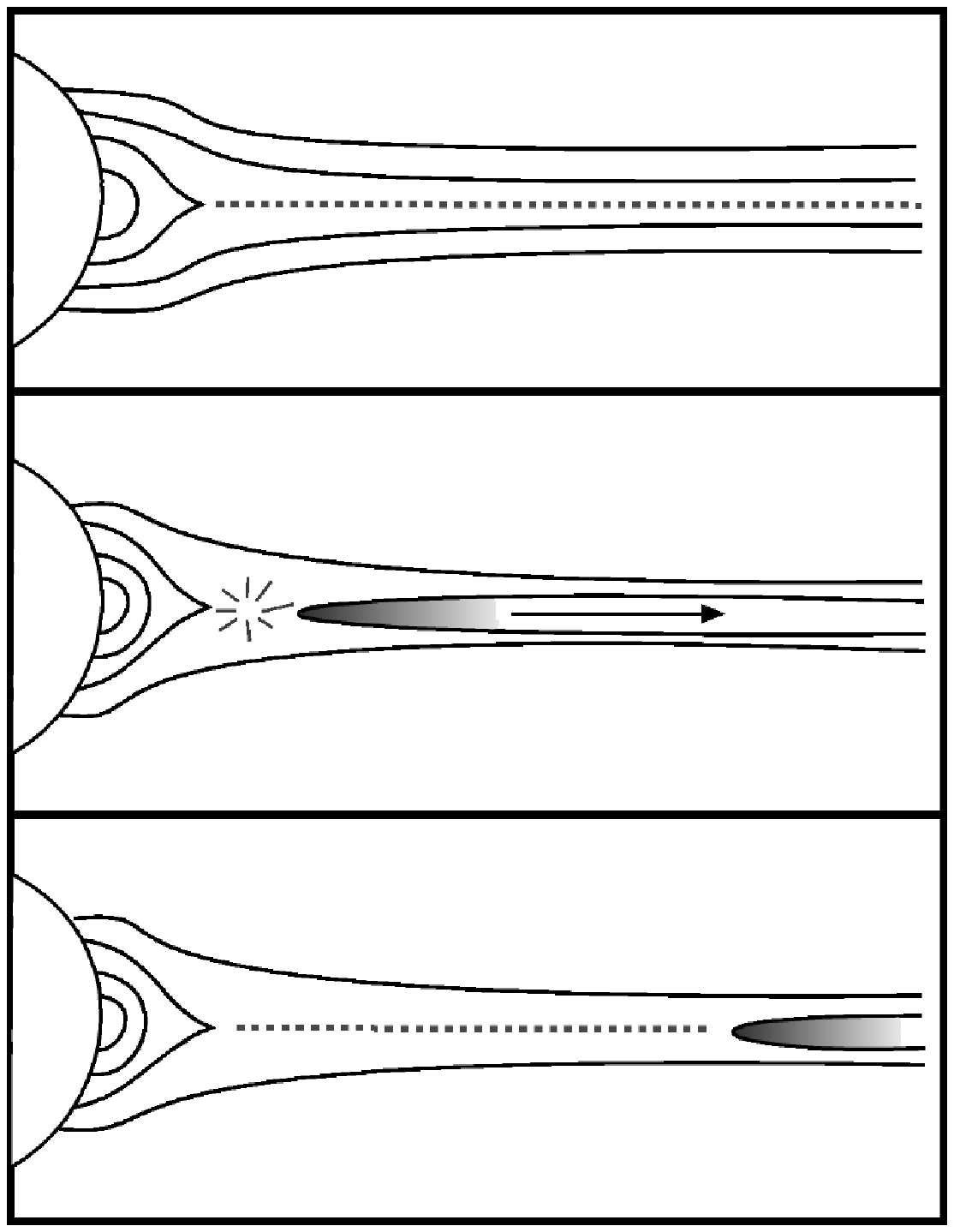}}
\end{center}

\noindent {\scriptsize Fig.\ 1. The formation of a blob. The top panel 
shows the initial situation: a  coronal streamer consisting of a helmet of 
closed loops surrounded by open field lines separated by a current 
sheet. In the lower two panels, reconnection has taken place between the 
two innermost field lines from the top panel. One new field line is a 
closed loop which becomes part of the helmet. The other one is now 
disconnected and moves outward, as shown in the bottom panels. On its 
way, it sweeps up plasma from the current sheet, which forms a blob.}
 
\newpage

Einaudi et al.\ \markcite{EEA98}(1998) and Dahlburg, Boncinelli, \& 
Einaudi \markcite{DBE98}(1998) performed numerical simulations of a 
current sheet, suggesting that the fast solar wind flowing along a 
current sheet can trigger, amongst others, a resistive instability, 
resulting in plasmoid formation. The initial situation in these 
simulations was a one-dimensional equilibrium,
a plane current sheet, which is then destabilized by two-dimensional 
disturbances. The model does not incorporate the helmet and the 
beginning of the current sheet, i.e.\ the cusp. We believe, however, that 
reconnection will occur preferentially just there. First, by virtue of 
the magnetic topology, the cusp is the most natural location for 
reconnection. Second, we notice that when plasma flows outward along the 
open field lines on the side of the helmet, an inward ``inertial 
pressure'' will arise at the cusp, where the plasma has to change 
direction to follow the field lines, which bend to become aligned with 
the streamer stalk. This pressure pushes the open field lines a little 
closer together right above the cusp. The reconnection that follows 
produces the new loops, as described above. Further study of both the 
location of reconnection and the subsequent movement of the blobs is 
needed. Numerical simulations would have to include, amongst others, the 
complete magnetic topology, the radially decreasing background density, 
and the acceleration of the surrounding solar wind.

%%%%%%%%%%%%%%%% EVIDENCE %%%%%%%%%%%%%%%%%%%%%

\section{Evidence}

Several features of the blobs support the proposed mechanism. The 
first observation is the acceleration pattern of the blobs. Sheeley
et al.\ \markcite{SEA97}(1997) observed that, in general, the blobs 
exhibit a fairly constant acceleration. However, in a speed-height plot 
representing measurements of about 65 independent blobs, they observed 
a peculiar ``corner'' at about $6$ or $7\,R_{\sun}$. Although 
this corner could be an artifact, Sheeley et al.\ \markcite{SEA97}(1997) 
remark that it might be a valid indication of a somewhat steeper initial 
acceleration. We suggest that the outward magnetic forces of the
newly formed field line which sweeps up the plasma could account
for this extra acceleration. As soon as $\beta > 1$ and the magnetic 
field no longer dominates the gas pressure, the field line just moves 
along with the slow solar wind, as Sheeley et al.\ \markcite{SEA97}(1997) 
suggested.

Second, the blobs often exhibit a concave-outward V-shape when the 
plasma sheet in which they move outward is slightly inclined to the line 
of sight \markcite{WEA98}(Wang et al.\ 1998). This supports the idea 
that a field line, with a V-(or U-)shaped trough at its bottom, is 
sweeping up plasma. However, the fact that the V-shape only becomes 
visible when the sheet is seen at an angle calls for a further 
explanation, which is illustrated in Fig.~2. This figure 
represents the blob creation, but now seen both from our viewpoint, in 
the plane of the sky (bottom row) and from the solar pole, in the plane 
of the current sheet (top row). The two field lines shown in the top 
left panel correspond to the innermost open field lines in the bottom 
left panel. The outgoing magnetic flux rope has little room to expand in 
the direction perpendicular to the current sheet, which, by its very 
nature, is very thin and bounded by magnetic fields. Thus, when the current 
sheet is seen side-on, the swept-up plasma looks like a narrow blob, as 
shown in the bottom right panel. However, there probably is a 
substantial azimuthal component to the magnetic field of the streamer 
(top left panel), e.g.\ due to differential rotation. This leads to a 
V-shape of the newly reconnected field line when seen from above. 
Consequently, from this point of view, the plasma that is swept up will 
be V-shaped as well (top right panel). As long as the current sheet is 
perpendicular to the plane of the sky, this does not affect our image of 
the blob; we simply see a projection that is still shaped like an 
elongated blob (as in the bottom right panel). However, when the sheet 
is inclined to the line of sight, the V-shape will become visible. We 
notice that an ESA Polar Orbiter at about 0.5 AU, a candidate mission 
for around 2007 \markcite{PEA98}(Priest et al.\ 1998), could observe the 
structures drawn in the top panels of Fig.~2.

\newpage

\begin{center}
\scalebox{0.57}{\includegraphics{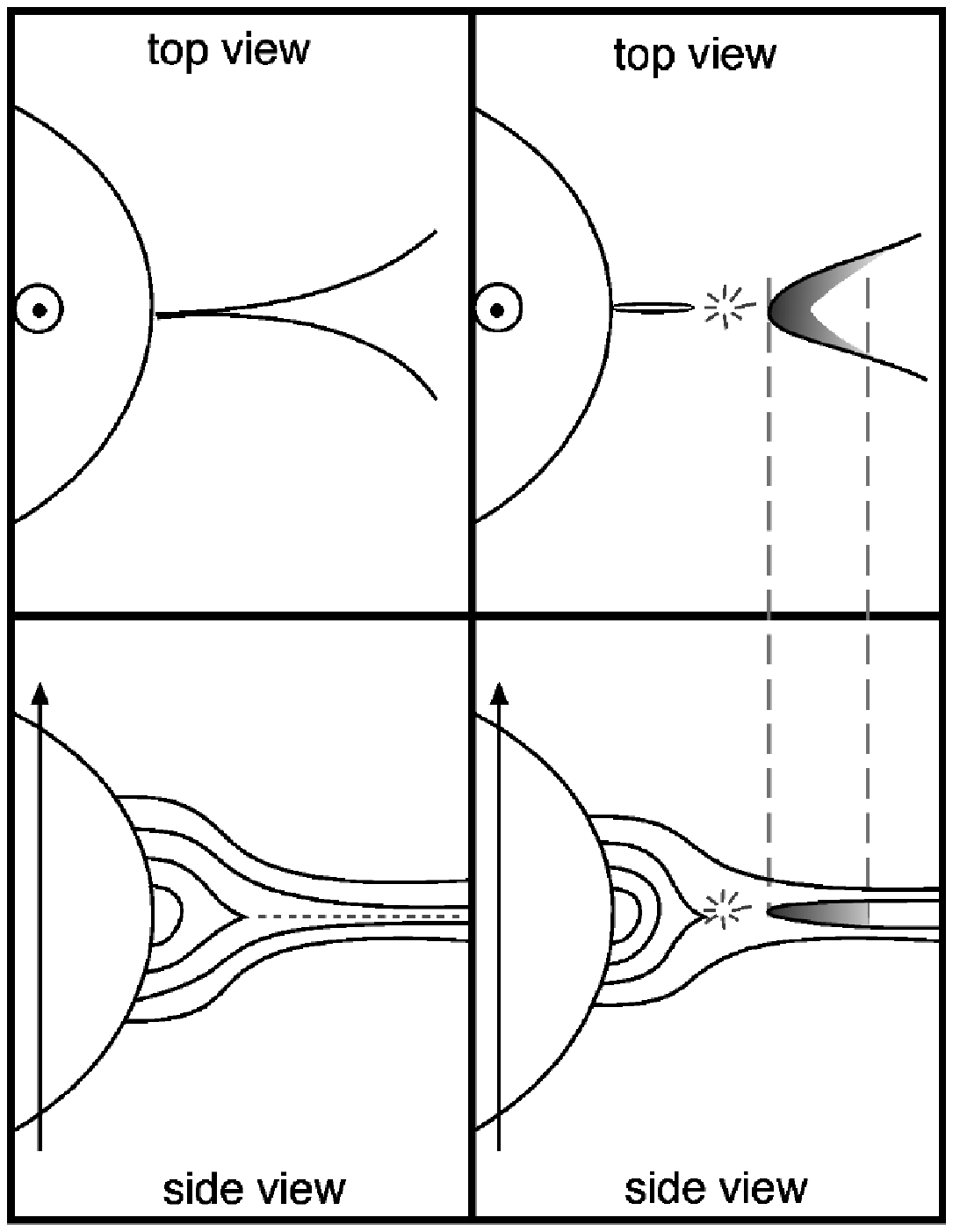}}
\end{center}

\noindent {\scriptsize Fig.~2. The formation of a blob, seen from two 
directions. The bottom panels show the streamer in the plane of the sky, 
as we see it. The top panels show the same picture, but now seen from 
above. The two field lines in the top left panel correspond to the 
innermost open field lines in the bottom left panel. The two left panels 
show the situation before, the two right panels after the reconnection.}

\vspace*{5 mm}

Third, it should be noted that the blobs remain coherent all through 
the LASCO-coronagraph's field of view, out to $30 \, R_{\sun}$. According
to Sheeley et al.\ \markcite{SEA97}(1997), they maintain constant 
angular spans and increase their lengths in rough accord with their 
speeds. This appearance is consistent with our model of a 
density enhancement that is bound on both sides by magnetic field lines,
which maintain a roughly constant angular span when they extend more or
less radially from the Sun (in the plane of the sky), and from below 
by the magnetic trough which initially sweeps up the material in the blob.

Fourth, the blobs are not very bright, but do contain quite some 
material right from the moment they appear above the helmet. Sweeping up 
the material in a field line that is moving outward in an exponentially 
decreasing density is a very effective way to accomplish that. Another 
way would be to collect the material in a closed field line in the 
helmet. Considering that there is no observational evidence for the 
complete eruption of such a closed field line, Wang et al.\ 
\markcite{WEA98}(1998) have suggested that the material is released 
along an open field line after a reconnection between a closed field 
line and a nearby open one. However, such a mechanism makes it much 
harder to explain the three observations mentioned above. Recent 
measurements of abundances in streamers by Raymond et al.\ 
\markcite{REA97}(1997) with the UVCS provide a definite test for whether 
the blob material comes out of the closed helmet or, as we propose, from 
the current sheet. These measurements show that elements with a high 
first ionization potential, like oxygen, are underabundant by an order 
of magnitude in the closed helmet, but only by a factor three on the 
sides of the helmet, where the field lines are open. Measurements of the 
abundances in blobs would thus allow us to determine whether their 
plasma originates in the closed helmet or in the open field lines 
adjoining the current sheet. In situ measurements of the slow solar wind 
provide another test: if the blobs are formed in the 
helmet instead of above it, the slow solar wind should contain a 
component exhibiting the specific characteristics of that 
material \markcite{WEA98}(Wang et al.\ 1998). 

\newpage

Fifth, Wang et al.\ \markcite{WEA98}(1998) observed that the creation 
of the  blobs is sometimes accompanied by a downflow of material. They 
compare this with downflows in the aftermath of CMEs, accompanying the 
closing down of fields blown open during the event. Clearly, this fits 
our model very well. While the outward moving loop will usually be most  
visible, the new closed loop will also collect some material while it is 
moving downward to become part of the helmet. 

Finally, we suggest that additional observational evidence for this 
mechanism can be collected by looking at the development of a helmet 
after the departure of a blob. The reconnection between open field lines 
not only releases a field line, which then forms a blob, but also adds a
new closed field line to the helmet, which should grow in size (with a 
little delay to allow it to be filled up with material from below). We 
notice that a measurement of this growth, combined with an estimate of 
the field strength at the bottom of the helmet, immediately yields the 
flux involved in the reconnection. Unfortunately, the solar rotation 
complicates this otherwise straightforward observational test. The 
additional field lines would presumably add at most a few percent to the 
size of the streamer. In coronagraph observations, such changes may just 
as well be the result of projection effects of structures that rotate 
into (or out of) view, aggravated by the contribution function for the 
visibility of material out of the plane of the sky 
\markcite{H93}(Hundhausen 1993). However, this problem can be overcome 
with a thorough statistical study of the relation between the expulsion 
of the blobs and the size and intensity of the helmet in a large number 
of streamers. We have not yet performed such a study, but preliminary 
observations of a smaller number of streamers yielded promising results. 
As an alternative to statistical studies, some of the problems of the 
solar rotation could also be overcome with the proposed NASA STEREO 
mission, to be flown from mid-2003 \markcite{R97}(Rust 1997), which 
would observe the Sun from two different angles.

%%%%%%%%%%%%%%%%% CONCLUSION %%%%%%%%%%%%%%%%%%

\section{Summary}

We have suggested that the origin of the blobs in coronal streamers 
that were observed by Sheeley et al.\ \markcite{SEA97}(1997) is related 
to the longstanding problem of maintaining a roughly constant flux in 
the interplanetary magnetic field. Two open field lines, from both sides 
of the current sheet, reconnect to form two new loops. One is a closed 
field line and becomes part of the helmet. The other is now disconnected 
from the Sun and moves outward, sweeping up the material that forms the 
blob. We have reviewed the observational evidence that support this 
idea: the acceleration pattern of the blobs, the V-shape that some of 
them exhibit, the collection of the material, the subsequent coherence of 
the blobs, and the observed retraction of the inner loop. Finally, we 
have suggested several observational tests for this theory. Numerical 
models should also be able to reproduce the proposed mechanism. 

%%%%%%%%%%%%%%%%% ACKNOWLEDGEMENTS %%%%%%%%%%%%%

\acknowledgements
{\it The authors thank S.\ Hill for his help in preparing the graphics,
J. Kuijpers for many discussions and helpful comments, and the
anonymous referee. M. K. van Aalst's work at the SOHO Experiment Analysis 
Facility at Goddard Space Flight Center was supported by ESA, the Olga 
Koningsfonds, the Leids Kerkhoven Bosscha Fonds, and Utrecht University. 
A. J. C. Beli\"{e}n carried out this work on an ESA Fellowship.}

%%%%%%%%%%%%%%%%% BIBLIOGRAPHY %%%%%%%%%%%%%%%%%

%%%%%%%%%%%%%%%%% END %%%%%%%%%%%%%%%%%%%%%%%%

\end{document}